\documentclass[manuscript,screen,nonacm]{acmart}
\AtBeginDocument{%
  }

\usepackage{multirow}
\usepackage{tabularx}
\usepackage{makecell}
\begin{document}

\title{Exploring a Design Framework for Children's Agency through Participatory Design}

\author{Boyin Yang}
\affiliation{%
  \institution{Department of Computer Science, University of Oxford}
  \city{Oxford}
  \country{United Kingdom}
}
\email{boyin.yang@cs.ox.ac.uk}

\author{Jun Zhao}
\affiliation{%
  \institution{Department of Computer Science, University of Oxford}
  \city{Oxford}
  \country{United Kingdom}
}
\email{jun.zhao@cs.ox.ac.uk}

\renewcommand{\shortauthors}{Yang et al.}


\begin{abstract}
Children’s agency plays a critical role in shaping children’s autonomy, participation, and well-being in their interactions with digital systems, particularly in emerging child–AI contexts. However, how designers currently understand and reason about children’s agency in practice remains underexplored. In this paper, we examine designers’ engagement with children’s agency through a participatory workshop in which we introduce a design-for-agency framework that supports designers externalising the consideration of agency in their design contexts. We find that while participants are committed to implementing ethical AI systems for children, they often struggle to understand why agency matters and how it can be operationalised in practice. Our agency design framework provided designers with a structured way to translate implicit, experience-based judgments into explicit articulations of agency trade-offs while acknowledging the associated design complexity. We conclude by offering initial insights into supporting designers’ reasoning about children’s agency and outlining directions for future research.
\end{abstract}




\maketitle



\section{Introduction}

Children’s agency in Artifial Intelligence (AI)-mediated systems (child-AI systems) has increasingly been recognised as a central value in the design of AI-enabled systems for children, when algorithmic manipulations and behavioural manipulations are prevalently applied to enable personalisation or so-called effective user engagement~\cite{kidron2018disrupted,unicef2021policy}. Prior work across human–computer interaction, child–computer interaction, and AI ethics has emphasised the importance of supporting children’s autonomy, participation, and meaningful involvement in systems that shape their learning, play, and everyday decision-making~\cite{druin1999role,rightsbydesign,iversen2017child}. As AI becomes more deeply embedded in children’s lives, agency is frequently raised as a key ethical and design principle, guiding discussions around responsibility, empowerment, and value-sensitive design.

However, despite this growing recognition, designing for children's agency remains challenging in practice. Agency is often articulated at a high level of abstraction, leaving design teams with limited guidance on how to translate such principles into concrete design decisions. 
In real-world design contexts, particularly in early-stage and exploratory phases, designing for agency and ethics is frequently treated as implicit, intuitive, or retrospectively justified, rather than explicitly reasoned at the start~\cite{mittelstadt2019principles,madaio2020co}. 

Existing toolkits, frameworks, and design guidelines have attempted to address this challenge by raising awareness or prompting reflection on ethical and social values~\cite{madaio2020co,d2021moral}. While many of these efforts have been positively received, particularly in their ability to sensitise designers to stakeholders perspectives~\cite{chivukula2024surveying}, they  have paid little attention to supporting the explicit consideration of agency for children. Furthermore, these tools largely focus on stimulating conceptual awareness of ethical designs rather than supporting design thinking: supporting designers’ reasoning processes, articulating implicit values, or shaping decision-making during design~\cite{friedman2019value,wilson2025towards}. This reasoning process is critical to translating abstract principles to concrete design practices systematically, which demands future research look beyond the usefulness of current tools and examine how to support designers’ reasoning about values, such as agency, during the design process.

In this paper, we hypothesise that designing for agency is fundamentally a cognitive and reasoning challenge, rather than merely a problem of optimising design outcomes or evaluating the effectiveness of specific design activities. This means that designing for agency should not be a tick-box exercise but involve a systematic process of recognising the value of supporting agency in a child-AI system, identifying children's needs for agency in relation to system structure and stakeholder roles, and reasoning about the trade-offs between designing for agency and other design priorities. From this perspective, the central question is not simply what agency means, but how designers think with agency during the design process: how they make sense of multiple forms of agency, prioritise among them, and reflect on their implications for system design.

To address this challenge, we introduce the Designing for Children’s Agency in AI (CHAI) framework, a conceptual framework designed to support designers’ reasoning about children’s agency in AI systems.
The framework distinguishes multiple types of agency and combines them with graded levels of support, encouraging designers to articulate and prioritise agency-related values explicitly. Rather than prescribing specific design solutions, the framework is intended to function as a cognitive scaffold: providing a structured way of externalising implicit assumptions, making value trade-offs visible, and situating agency considerations within concrete system contexts. 


To examine how the CHAI framework supports designers’ reasoning about children’s agency, we ran an exploratory study involving a series of participant workshops with early-career designers of child-AI systems to ask specifically:
\begin{enumerate}
    \item How do designers currently understand and engage with children’s agency when designing child–AI systems?
    
    \item How does a framework that fosters explicit considerations of agency affect designers' reasoning about agency?

    \item What challenges arise in applying the CHAI framework in practice?
\end{enumerate}

We report three key findings. First, while designers demonstrate an implicit awareness of children’s agency, this understanding is rarely articulated explicitly during design. Second, engaging with the CHAI framework enables designers to externalise agency-related considerations, shifting designers from intuitive, value-based judgments toward more explicit articulation and justification of agency. Finally, participants reported that although explicitly considering agency adds complexity to design, it leads to more nuanced and multidimensional reasoning about agency. Together, these findings demonstrate that our framework has the potential to function as cognitive support for agency-oriented design reasoning, rather than as a prescriptive tool for improving design outcomes.



\section{Related Work}

\subsection{Agency in Children's AI Systems}
\label{section:agency_in_children_ai_systems}

AI systems increasingly shape how children learn, socialise, and make decisions, raising concerns about long-term impacts on autonomy, well-being, and development~\cite{law2022examining,lin2021parental,amnesty2023}. While much prior work focuses on safeguarding children from harm, less attention has been paid to how design choices affect children’s ability to act meaningfully within AI-mediated systems. This is particularly problematic for vulnerable children, whose developing cognitive and socio-emotional capacities make them more susceptible to persuasive or manipulative interactions~\cite{im2025}.

Agency is closely related to the concept of autonomy. \textit{Autonomy} is generally defined as the capacity to act in alignment with one’s values and intentions~\cite{prunkl2022human}. In psychology, agency is more broadly understood as \textit{the ability to make meaningful choices and act upon them}~\cite{bandura2001social}. Importantly, agency is not limited to individual control (i.e., \textit{individual agency}) but includes relational forms such as \textit{proxy agency}, which occur when individuals rely on others who hold resources, expertise, or power to act on their behalf; \textit{co-agency}, which refers to cases where an individual’s ability to act is co-developed with or supported by others; and \textit{collective agency},  which arises when people work together to achieve shared goals that none could accomplish alone~\cite{bandura2001social}. For children, whose decision-making is inherently shaped by caregivers, peers, and institutions, these relational dimensions are particularly salient.

Designing for children's agency in the context of AI systems is particularly relevant considering the prevalent use of algorithmic manipulation and behaviour engineering in large AI platforms~\cite{5rights2021pathways}. However, despite its importance, children’s agency is often treated implicitly in AI design, either assumed to be weak and therefore overridden by protective measures, or presumed to be present without careful consideration of developmental appropriateness. This lack of conceptual clarity risks both \textit{over-protecting} children in ways that constrain learning and autonomy, and \textit{overlooking} vulnerabilities that require support.

\subsection{Children's Agency and Ethical AI Design}

Designing for children’s agency is closely intertwined with broader ethical concerns in AI, including transparency, accountability, fairness, privacy, and safety~\cite{wang_challenges_2024}. Prior reviews of child-focused AI systems and policies highlight significant gaps in how these ethical principles are addressed, with limited attention to children’s specific needs and roles~\cite{wang2022_informing,wang_challenges_2024}. While protection remains a dominant framing in ethical AI discourse, an exclusive focus on safeguarding risks marginalising children’s participation and voice. Agency should be treated not as an optional ethical principle but as a foundational lens through which other ethics principles are realised in practice~\cite{zhao2025}. Without agency, transparency cannot be meaningfully interpreted, accountability cannot be exercised, and fairness cannot be contested. Designing for agency thus operationalises ethical commitments by enabling children to understand, question, and engage with AI systems in developmentally appropriate ways~\cite{el2024insight}.

\subsection{Ethics and Agency by Design}

The idea of ethics by design promotes the systematic integration of ethical considerations into the design and development of AI systems. While the approach has gained prominence in research and policy~\cite{brey2024ethics}, its practical application remains challenging due to limited actionable guidance for practitioners~\cite{peterson2024embedding}. The ``Ethics by Design for AI'' (EbD-AI) framework~\cite{brey2024ethics} addresses this gap by providing a structured process for considering moral principles such as human agency, privacy, fairness, openness, and well-being throughout system development. EbD-AI aligns ethical reflection with development stages, such as assessment, mapping, and implementation, while recognising that the concrete realisation of these tasks must remain context-dependent.

Beyond embedded ethics frameworks, a range of approaches aim to promote responsible AI, including autonomous ethical agents and ecosystem-level perspectives~\cite{cervantes2016autonomous,tiribelli2024rethinking}. However, these approaches face persistent challenges in operationalising abstract ethical principles within situated socio-technical contexts, particularly when ethical concerns must be negotiated across multiple stakeholders and system components~\cite{stahl2024ethics}. As a result, translating ethical principles into concrete design decisions remains difficult in practice.

Within this landscape, supporting children’s agency has emerged as a critical design concern. Child-centred research has proposed several mechanisms for fostering children’s digital autonomy, including contextualisation, nudging, peer support, and scaffolding~\cite{wang2023empower}. 
While these mechanisms address important aspects of children’s cognitive, motivational, and self-regulatory development, they primarily focus on \textit{individual} agency. Comparatively little attention has been given to the relational dimensions of agency, namely how children’s actions and decisions are shaped by caregivers, peers, technologies, and different social contexts. The Responsible Innovation in Technology for Children (RITEC) Digital Design toolbox from LEGO and UNICEF, developed through deep research with children, supports designers incorporating considerations of children’s well-being in the design process through an explicit value reasoning process~\cite{unicef2024ritec}. 
This toolkit foregrounds children’s well-being and social connections in play-based design, which provides an inspiration for our framework, which focuses on designing for children's agency development.


\section{The CHAI Framework}

The Designing for Children’s Agency in AI (CHAI) framework provides a conceptual structure to support designers’ reasoning about children’s agency in the context of AI-enabled systems. It draws inspiration from established approaches to ethical and child-centred technology design, including the EbD-AI framework~\cite{brey2024ethics} and RITEC design toolbox~\cite{unicef2024ritec}. It also aligns with rights-by-design~\cite{livingstone2023child} and playful-by-design~\cite{colvert2024playful} principles to prioritise children's rights and creativity. While these approaches provide important normative and procedural guidance, CHAI focuses specifically on supporting designers’ reasoning processes by offering a deliberately simple yet structured way to engage with children's agency during design. In doing so, the framework bridges abstract principles with actionable design practices, while remaining adaptable to different design contexts. 

The framework was refined through three iterative pilot workshops with a child–AI design competition winner team, informing both its conceptual structure and integration with concrete design contexts.

\subsection{A Four-Stage Reasoning Process}
\begin{figure}[ht]
    \centering
    \includegraphics[width = \textwidth]{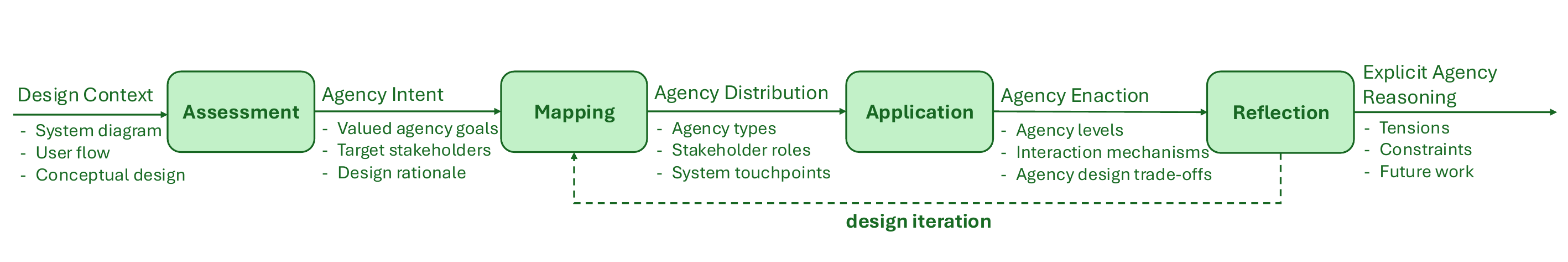}
    \caption{The four-stage reasoning process of the CHAI framework, showing how designers’ reasoning about children’s agency is transformed from an initial design context into explicit agency design through assessment, mapping, application, and reflection. The process produces intermediate reasoning artefacts and allows iteration among later stages. }
    \Description{The four-stage reasoning process of the CHAI framework, showing how designers’ reasoning about children’s agency is transformed from an initial design context into explicit agency design through assessment, mapping, application, and reflection. The process produces intermediate reasoning artefacts and allows iteration among later stages.}
    \label{fig:four_stages}
\end{figure}

Inspired by the highly effective EbD-AI toolkit to support ethical designs~\cite{brey2024ethics}, the core of the CHAI framework is a four-stage process for supporting designers’ reasoning about children’s agency, including assessment, mapping, application, and reflection.  The reasoning process captures the core cognitive mechanisms that designers can draw on by using the types and levels of agency as the reasoning lenses, to externalise values, structure reflection, and surface practical challenges in applying design-for-agency principles (see Figure~\ref{fig:four_stages}). 

The \textbf{assessment} phase initiates a design team's reasoning of children’s agency in their existing design contexts to articulate their agency-related design goals, clarify which forms of agency are valued, for whom, and why. The goal is to encourages them to progressively make agency-related assumptions and goals explicit and move designers towards a shared and articulated understanding of intended agency design goals within the system.

The \textbf{mapping} phase supports structured reasoning about how agency considerations relate to different parts of a child-AI system. 
This helps the design team to externalise agency types and levels that are meaningful to them by linking abstract values to concrete system components and interactions that contribute to the overall agency goal identified in Phase~1.

The \textbf{application} phase focuses on enabling agency in practice. Designers can take the key system components related to agency and consider whether they need adjust their current design by introducing new functions or supporting interactions with new types of stakeholders.

The \textbf{reflection} phase provides space for designers to examine how their reasoning about children’s agency evolves through engaging with the CHAI framework, to surface tensions, limitations, and breakdowns encountered in practice with their team. Rather than introducing new design activities, this phase focuses on looking back at the reasoning process and its outcomes and prepares for actual implementations of redesigning.

\subsection{Four Types of Agency as Reasoning Lenses}
To assist designers to externalise the types of agency relevant to their systems and stakeholders at the assessment and mapping stages, the CHAI framework introduces four analytically distinct types of agency (introduced in Section~\ref{section:agency_in_children_ai_systems}) that function as reasoning lenses during design.

\begin{figure}[ht]
    \centering
    \includegraphics[width = 0.6\textwidth]{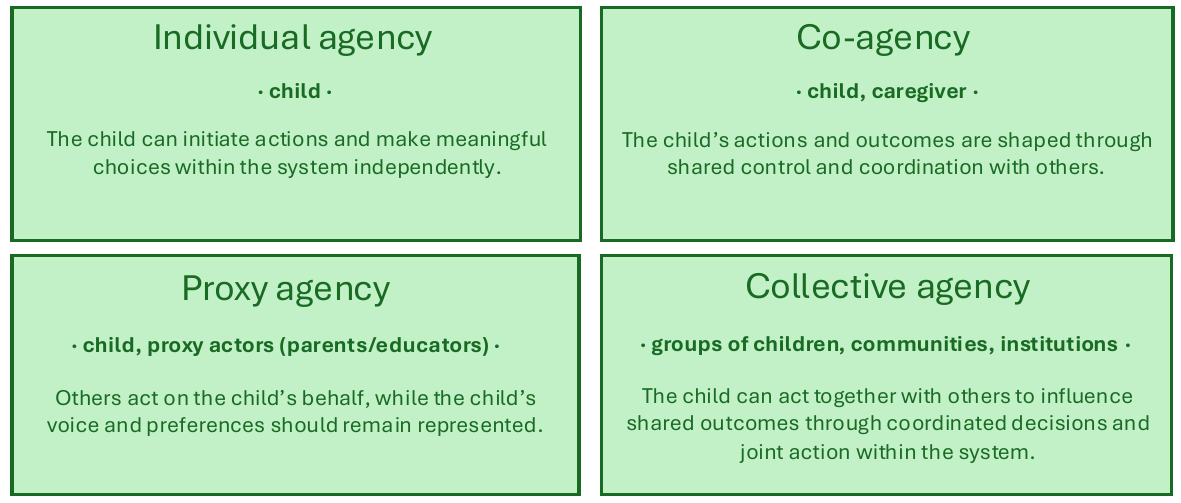}
    \caption{Four types of agency used in the CHAI framework as reasoning lenses. The visual abstractions represent different loci of action, control, and coordination across key stakeholders in each agency type. This illustration is used as part of the workshop materials to support designers’ reasoning about types of agency during the design process.}
    \Description{Four types of agency used in the CHAI framework as reasoning lenses. The visual abstractions represent different loci of action, control, and coordination across key stakeholders in each agency type. This illustration is used as part of the workshop materials to support designers’ reasoning about types of agency during the design process.}
    \label{fig:types_of_agency}
\end{figure}

These lenses provide lightweight prompts for considering how children’s agency is distributed across stakeholders and system components.
This explicit conceptualisation of agency is drawn from a literature review of agency conceptualisation from multiple disciplines, including social psychology, education, and philosophy. 
As shown in Figure~\ref{fig:types_of_agency}, the four types of agency function as distinct reasoning lenses, including individual action, shared coordination, mediated action through others, and group-level collective action. The social and relational dimensions of agency are primarily informed by Bandura’s theorisation, which positions agency as inherently shaped by social contexts and interpersonal dynamics~\cite{bandura2001social,bandura2000exercise}.

\begin{description}
    \item[Individual agency] refers to a child’s capacity to make choices and act independently within or through the system. This encourages designers to identify where children can initiate actions on their own, make meaningful selections, or exercise control over outcomes, and where system constraints (e.g., defaults, automation, friction, or permissions) may limit such autonomy.
    \item[Co-agency] refers to situations where two or more individuals \textit{share} control, responsibility, and influence over a task, outcome, or process. This encourages designers to examine how agency is negotiated in child–adult or child–peer interactions: for example, how a parent, teacher, or peer may guide, shape, or override a child’s actions, and how the system mediates these shared decision processes. 
    \item[Proxy agency] describes situations where a child \textit{relies on} others to act on their behalf to secure desired outcomes. This focuses designers' attention on mediated interactions; for instance, when parents or educators configure, approve, or trigger actions for a child, or when automated system functions act for the child. Crucially, the proxy lens foregrounds the need to respect children’s voice within mediated arrangements, prompting designers to consider how children’s preferences are represented, how they can contest or revise proxy decisions, and how accountability is maintained when others act on their behalf. 
    \item[Collective agency] is exercised through socially coordinative and interdependent effort, when people act together to influence outcomes they value by pooling resources, actions, and emotional support. This encourages designers to move beyond individual interactions and consider how groups, such as classmates, families, peer communities, or broader support networks, jointly shape outcomes, and how the system enables (or constrains) collective participation, coordination, and shared responsibility.
\end{description}

\subsection{An Agency Mapping Matrix as Value Reflection Lenses}
Building on the four types of agency, the CHAI framework also comes with an agency mapping matrix, consisting of three levels of agency (low, medium, and high) along with four types of agency, as value-reflection lenses. These lenses encourage designers, during the \textit{mapping} stage, to make explicit how each type of agency is expected to operate within their child–AI system, while considering children’s rights and voices. Inspired by the RITEC toolkit~\cite{unicef2024ritec}, the three levels provide a structured way for designers to reason about degrees of participation, control, and autonomy as experienced by the child, either independently or through interactions with different types of others.

\begin{figure}[ht]
    \centering
    \includegraphics[width = 0.8\textwidth]{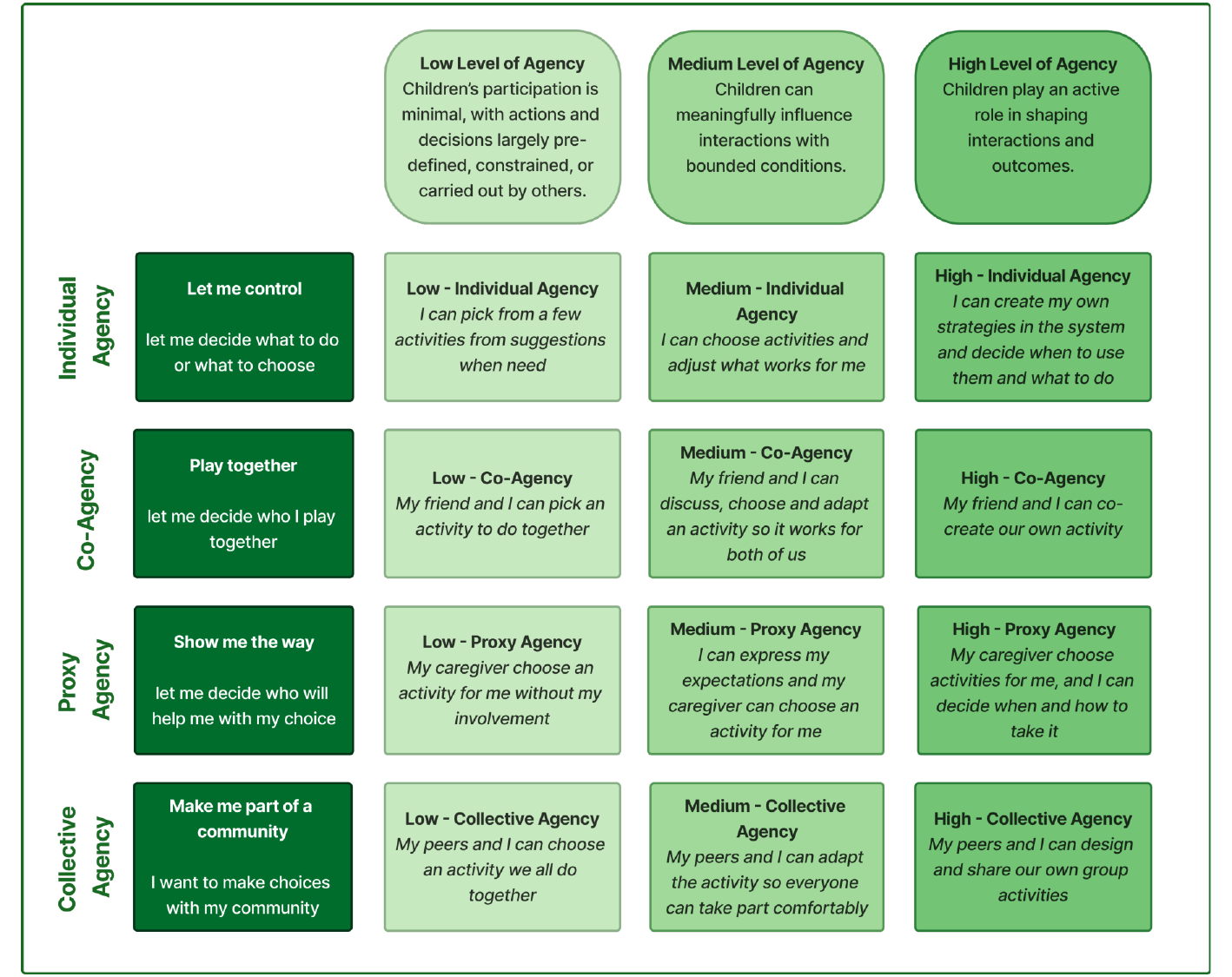}
    \caption{Three levels of agency used in the CHAI framework as reasoning lenses. The levels capture differences in children’s involvement in initiating actions, influencing decisions, and shaping outcomes, as well as how these processes are mediated by other stakeholders and the AI system. This figure is used as part of the workshop materials in this study.}
    \Description{Three levels of agency used in the CHAI framework as reasoning lenses. The levels capture differences in children’s involvement in initiating actions, influencing decisions, and shaping outcomes, as well as how these processes are mediated by other stakeholders and the AI system. This figure is used as part of the workshop materials in this study.}
    \label{fig:levels_of_agency}
\end{figure}

Figure~\ref{fig:levels_of_agency} illustrates how a sense of agency could be described from the child’s point of view, reflecting what it feels like to have a particular degree of control or involvement in an interaction. 
The statements provided by the mapping matrix serve as concrete prompts that help designers imagine how different configurations of control, autonomy, and responsibility may be experienced by children in specific design scenarios.

At a \textbf{low level of agency}, children’s participation is minimal, with actions and decisions largely predefined, constrained, or carried out by others (e.g. systems or adults). The child’s role is primarily reactive, with limited opportunities to influence how interactions unfold. This level prompts designers to reflect on situations where agency may be intentionally or accidentally restricted, delegated, or scaffolded, and to consider the implications for children’s sense of control and voice.

At a \textbf{medium level of agency}, children can meaningfully influence interactions within bounded conditions. Agency at this level is characterised by choice, adjustment, negotiation, or shared decision-making, often involving coordination with designed AI functions. The medium level foregrounds partial autonomy and collaborative control, encouraging designers to think about how systems can support dialogue, adaptation, and mutual understanding when supporting full independence is either infeasible or inappropriate.

At a \textbf{high level of agency}, children play an active role in shaping interactions and outcomes. They may initiate actions, develop strategies, or co-create activities with the AI system, exercising substantial control over how the system is used. This level encourages designers to consider how AI systems can support proactive participation and intentional action, while remaining attentive to social, technical, and institutional constraints.

Degrees of support for agency do not always map directly onto children’s developmental maturity. For example, a system may provide a high level of individual agency for children under five by allowing them to choose their reading programmes. We also emphasise to designers that maximising support for all types of agency is not necessarily an optimal design goal, as tensions and trade-offs often arise when supporting different forms of agency. Instead, children’s voices and best interests should guide design decisions.

\section{Study Design and Methods}

\subsection{Workshop Structure}
To examine how the CHAI framework influences designers’ reasoning about agency during design, we conducted participatory workshops with designers building AI systems for children. In each phase of the workshop, a researcher first introduced the task and explained the relevant elements of the CHAI framework, then facilitated design teams as they independently applied the framework to their design thinking. This approach allowed us to observe how teams used the CHAI framework on their own and to identify what additional support might be needed to make the framework a self-contained resource for designers. Table 1 provides an overview of the workshop structure.

\begin{table}[h]
\begin{tabular}{llllll}
\hline
\textbf{Phase}      & \textbf{No.} & \textbf{Activity step}          & \textbf{RQ} & \textbf{CHAI Stage} & \textbf{Data collected}                   \\ \hline
\multirow{3}{*}{1} & 1            & Design Recap           & RQ1         & --                  & Screen + audio recording        \\
                    & 2            & System Visualisation   & RQ1         & --                  & Screen + audio recording        \\
                    & 3            & Pre-agency Elicitation & RQ1         & --                  & Questionnaire + audio recording \\ \hline
\multirow{5}{*}{2} & 4            & CHAI Introduction      & --          & --                  & Screen + audio recording        \\
                    & 5            & Agency Intent          & RQ2         & Assessment          & Screen + audio recording        \\
                    & 6            & Agency Annotation      & RQ2         & Mapping             & Screen + audio recording        \\
                    & 7            & Critical Redesign      & RQ2/RQ3     & Application         & Screen + audio recording        \\
                    & 8            & Reflection Interview   & RQ3         & Reflection          & Audio recording                 \\ \hline
3                  & 9            & Post-questionnaire     & RQ2/RQ3     & --                  & Questionnaire                   \\ \hline
\end{tabular}
\caption{An overview of the workshop structure, showing each phase, activity step, associated research questions (RQ), corresponding CHAI stages, and data collected.}
\label{tab:workshop_structure}
\end{table}

\subsubsection{Phase 1: Introduction}
We began with a five-minute introduction outlining the goals and structure of the participatory workshop. All participants provided informed consent to take part in the study, including permission for audio and video recording and for their design artefacts to be collected via Figma boards. Participants then completed a pre-workshop questionnaire about their prior experience designing for children and their understandings of agency. We emphasised that there were no correct answers and that the workshop was not intended to evaluate or judge participants’ existing design practices.

\subsubsection{Phase 2: Main session}
The main session was structured according to the reasoning process of the CHAI framework and consisted of four phases: assessment, mapping, application, and reflection. Each phase comprised a small set of design activities that progressively supported designers in articulating, structuring, enacting, and reflecting on children’s agency in their systems. At each phase, a researcher introduced the task and the relevant elements of the CHAI framework and facilitated discussion, while design teams applied the framework independently. All teams were familiar with using Figma and documented their design outcomes for each phase on shared boards. Researchers primarily served as facilitators of the activities and discussions.

\textit{Phase 2.1 Assessment -- Design Context and Agency Intent.}
\label{section:phase_1_assessment} Designers first recapped their existing projects by articulating primary design goals and externalising their designs through system diagrams and user scenarios. They articulated their initial understandings of agency and confidence in designing for agency, providing a baseline for subsequent reasoning. 
The facilitator then introduced the four types and three levels of agency as the reasoning lenses for designers to consider how agency may be enabled, constrained, or redistributed in child–AI systems. Designers applied these lenses to their own projects to articulate agency-related design goals. The outcome of this phase was a shared articulation of design objectives and the intended role of children’s agency. 

\begin{figure}[ht]
    \centering
    \includegraphics[width = 0.8\textwidth]{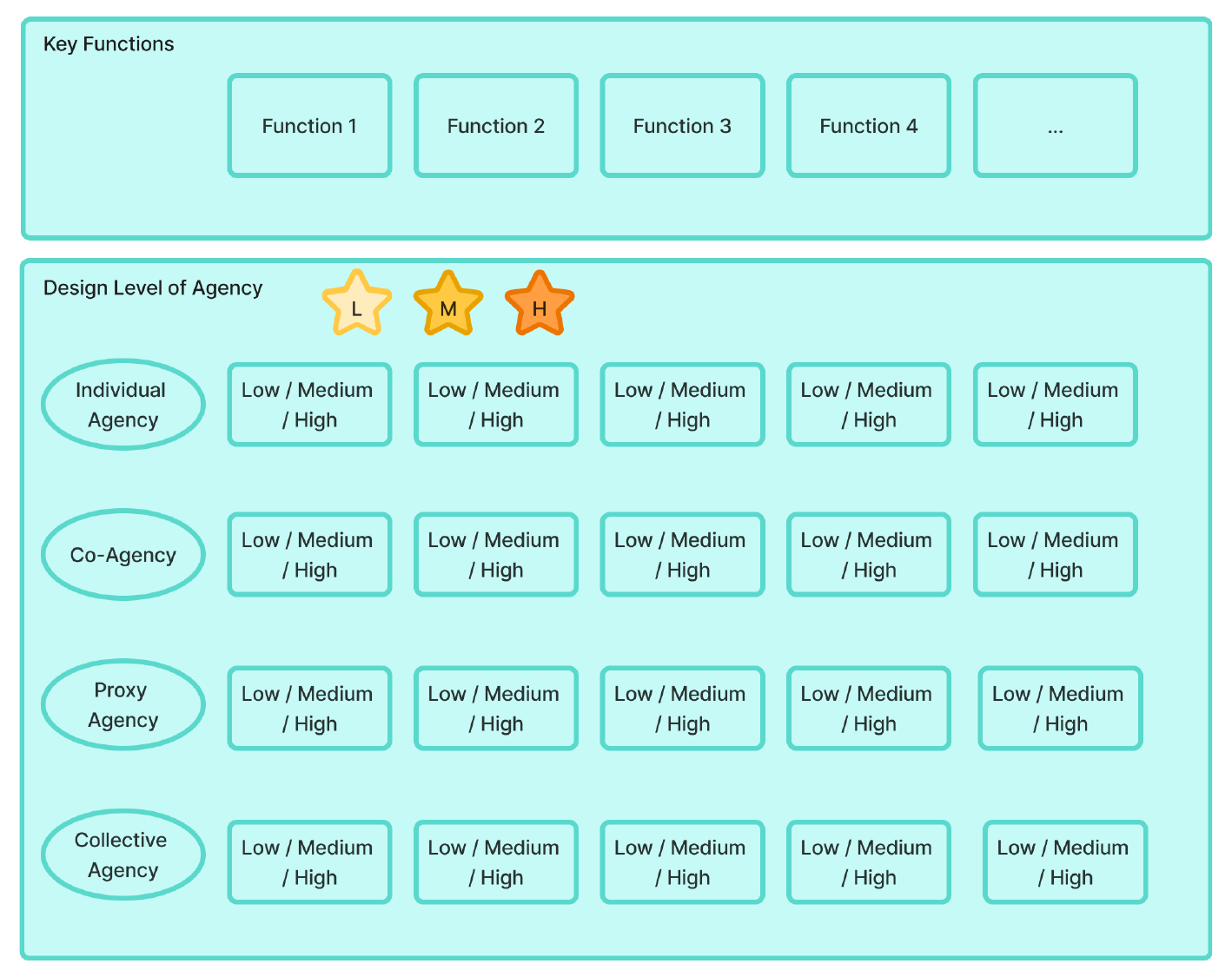}
    \caption{Agency mapping matrix used in the CHAI participatory design workshop. Designers place key system functions along the top row and annotate each function with intended levels of support (low, medium, high) across four types of agency: individual, co-, proxy, and collective.}
    \Description{Agency mapping matrix used in the CHAI participatory design workshop. Designers place key system functions along the top row and annotate each function with intended levels of support (low, medium, high) across four types of agency: individual, co-, proxy, and collective.}
    \label{fig:mapping}
\end{figure}

\textit{Phase 2.2 Mapping -- Agency Annotation.}
Using the agency mapping matrix (Fig.~\ref{fig:mapping}), designers examined their system component to 1) identify where agency was relevant in their system, and ii) explore appropriate degrees of participation, control, and autonomy from the child’s perspective. Agency levels served as prompts for reflecting on interaction dynamics and trade-offs between stakeholders and the system. This stage produced an agency priority map that externalised agency types and levels for each team's design.

\textit{Phase 2.3: Application -- Critical Redesign.}
In the application phase, designers reflected on how well their current designs supported the target levels of agency, identified in the mapping phase. Each team selected a system function that was most consequential for their agency design goals and examined it in greater depth. Designers explored alternative design choices and proposed redesigns to better support children’s agency, while remaining aligned with the original design goals and considering relevant constraints and trade-offs.

\textit{Phase 2.4: Reflection -- Reasoning Review.}
Designers reflected on how their reasoning about children’s agency evolved through engaging with the CHAI framework. They shared design tensions, trade-offs, and alignments between agency objectives and initial design goals through group discussions. This provides the critical inputs for potential follow-up design work or implementation. 

\subsubsection{Phase 3: Wrap-up and Feedback}

The workshop concluded with a semi-structured interview examining how the agency-oriented reasoning shaped participants' thinking and how agency-related goals conflicted with other design constraints. These discussions foregrounded challenges in operationalising the CHAI framework and highlighted areas where the framework felt difficult or incomplete. Participants also completed a post-workshop questionnaire matched to the pre-workshop measures, with additional items assessing perceived usefulness and difficulty of applying the CHAI framework. Together, these reflective activities supported analysis of changes in designers’ reasoning and surfaced practical boundaries and limitations of the framework.

\subsection{Participants}
Participants were recruited as small design teams (1--4 participants per group) who had previously taken part in a child–AI system design competition or hackathon. All participants entered the study with an existing child-focused AI system design project, which served as the design context for the workshop activities. In total we had four teams and nine designers participating in our workshops, as summarised in Table 2, with T0 participating both in our pilot studies and the participatory workshop. Participants included undergraduate, Master’s, and doctoral students with backgrounds in education, design, engineering, or related fields. While the pilot sessions informed the refinement of the workshop design, the analysis in this paper primarily draws on data from the participant workshops. All participants provided informed consent prior to participation. Workshops were conducted online using video conferencing tools and were audio- and screen-recorded for analysis.


\begin{table}[h]
\centering
\begin{tabularx}{\linewidth}{lllllll}
\hline
\textbf{Team}       & \textbf{ID} & \textbf{Role}         & \makecell[l]{\textbf{Design} \\ \textbf{Experience}} & \makecell[l]{\textbf{Child-Centred} \\ \textbf{Project Experience}}             & \makecell[l]{\textbf{AI Design} \\ \textbf{Familiarity}} & \makecell[l]{\textbf{Agency / Ethics} \\ \textbf{Experience}}     \\ \hline
\multirow{4}{*}{T0} & P1          & Master's student      & $\leq$ 1 year           & Limited experience  & Not familiar                   & No \\
                    & P2          & Master's student      & $\leq$ 1 year           & Limited experience  & Not familiar                   & No \\
                    & P3          & Master's student      & 1--2 years                 & Limited experience  & Slightly famillar              & No \\
                    & P4          & Master's student      & $\leq$ 1 year           & Limited experience  & Not familiar                   & No \\ \hline
\multirow{2}{*}{T1} & P5          & Undergraduate student & $\leq$ 1 year           & Limited experience  & Slightly famillar              & No                                              \\
                    & P6          & Master's student      & 1--2 years                 & Moderate experience                & Moderately famillar            & Yes \\ \hline
\multirow{2}{*}{T2} & P7          & PhD student           & $\leq$ 1 year           & Moderate experience                & Not familiar                   & Yes \\
                    & P8          & Master's student      & $\leq$ 1 year           & Limited experience  & Slightly famillar              & No                                              \\ \hline
T3                  & P9          & Master's student      & $\leq$ 1 year           & Limited experience  & Not familiar                   & No                                              \\ \hline
\end{tabularx}
\caption{Overview of participant roles, design experience, and prior exposure to child-centred and AI ethics frameworks.}
\label{tab:participants}
\end{table}

%

Each workshop lasts approximately 60--90 minutes and is conducted remotely using Microsoft Teams. Collaborative design activities are supported through shared screens and a digital whiteboard (Figma), which participants use to create and modify system diagrams, annotate agency-related considerations, and explore redesign ideas in real time. All workshops follow the same overall structure described above.


\subsection{Data Analysis}
Workshop sessions were audio- and screen-recorded and transcribed for analysis informed by reflexive thematic analysis~\cite{braun2006using}. Both authors coded the transcripts after each workshop and met to discuss interpretations and reflect on the evolving codebook, which was iteratively updated across workshops. During the process, we focused on identifying recurring patterns, with attention to participants’ baseline consideration of agency, moments of agency-related reasoning (e.g., identification of agency, system functions, stakeholder and interaction changes), and reflections on challenges, complexities, and trade-offs. We conducted a final review of the codebook by aligning our understanding of code definitions and resolving differences through discussion. Codes were grouped into higher-level themes capturing how the workshop supported or challenged participants’ reasoning about agency. The analysis aimed to characterise patterns in designers’ reasoning rather than evaluate learning outcomes or compare performance across groups.
Data were primarily analysed at the team level (T0--T3); individual participants (e.g., P1, P2) are referenced only when notable differences emerged within a team.


\section{Findings}
\subsection{Agency Was Largely Absent or Implicit in Designers’ Initial Reasoning}

Prior to engaging with the CHAI framework, designers shared that they rarely articulated children’s agency as a distinct design concern. Although participants described their designs were aimed to be child-centred, safe, or supportive, agency-related considerations, such as who initiates actions, who makes decisions, and whose voice is represented, were often embedded implicitly within functional or safety-driven choices, or even missing. 

Designers often frame their systems primarily around parents or caregivers, with children positioned as indirect beneficiaries. For example, T1 shared that their project focuses on supporting parents logging meltdown data, while treating children as subjects of recording rather than active users. 
T3 explicitly stated that ``parents'' are their target audience, positioning the child as someone to be considered within a system, rather than an actor with decision-making power.

\subsection{The CHAI Framework Enabled Explicit, Function-Level Reasoning About Agency}
The CHAI framework helped participants map agency types and levels onto specific system functions, translating abstract principles into specific design opportunities. This shift was consistently observed across T1--T3.

Using system diagrams as a shared reference point, designers demonstrated an ability to identify and annotate specific system functions with different agency configurations. 
For example, T3 reasoned that collecting children’s narratives required high individual agency, while generating personalised stories involves high co-agency between children and parents, and shared reading activities are framed as medium-level proxy agency guided by parent and mediated by AI. 

This function-level analysis prompted designers to articulate why certain agency configurations were appropriate in particular contexts. Participants debated whether parental involvement constituted co-agency or proxy agency, and whether children’s input meaningfully influenced outcomes or merely served as data for automated processes. Notably, T0 and T1 reflected that the framework provided ``a shared vocabulary'' that made such reasoning discussable and communicable. 

\subsection{Making Agency Explicit Identifies Misalignments Between Stated Values and Design Decisions}
As designers map agency across system functions, previously unexamined tensions between design intentions and actual design decisions become visible. Designers began to recognise unintentional exclusion of children and the tension of imposing `adult shoes' in their designs.

Across multiple workshops, participants had often justified limiting children’s involvement by discussing their priorities of ensuring safety, efficiency, or developmental appropriateness. By applying the CHAI framework to their systems, participants demonstrated a \textit{recognition of their unintended exclusion of the voice of children} in their design. For example, T1 initially focused on creating a ``reflective support system'' for the meltdowns of neurodivergent children and explicitly stated that the app was ``about direct interaction between children''. However, during the workshop, they recognised that their system was designed to  limit ``screen time'', which ensured data was used for clinical decisions but simultaneously limited child interaction.
Similarly, while the stated goal of T3 was to help children understand emotions, the designer admitted that their perception of ``child is not our target'' removed the child from the primary interaction loop.

The framework enabled designers to identify explicitly who holds decision-making power and exposed the ``adult shoes'' tension. For example, P9 noted that while ``structured questions'' enable participation from children with limited verbal skills, they also brought a possibility of overlooking agency of choice. Reflection led P5 to realise that children might feel ``frightened'' by being recorded without consent during their interactions with their system, a tension between the ``parent's right to record'' and the ``child's right to privacy'', and thus revised their original design.

\subsection{Making Agency Explicit Surfaced Trade-offs and Increased Perceived Design Complexity}
Participants from all sessions (T0--T3) described the process of designing for agency as making their designs ``more complicated'' or ``heavier,'' particularly as they recognised that single features often entailed multiple, overlapping forms of agency distributed across children, parents, and the AI system. However, they also recognise that this complexity is a necessary consequence and encourages structure design thinking.

Participants characterised this increased complexity as a meaningful consequence of treating agency systematically rather than as a general property. For example, P5 raised that involving more stakeholders made the process ``a bit more messy'' but the resulting app would be ``better in the end''. P8 noted that thinking through both sides was ``definitely more complicated,'' while P7 further noted that this complexity provided a ``more comprehensive picture of the user journey''.

This shift towards externalising the value of agency encouraged divergent design thinking during early stages where value commitments are often left implicit. As T1 reported, the framework opened up ``another branch'' of development they had not previously considered, such as the child's right to understand and consent to data usage. T2 described the framework as providing ``positive constraints,'' preventing ideas from ``flying everywhere'' and forcing deeper reflection on user experience. Furthermore, T3 reflected that they moved from ``abstract'' theoretical knowledge to ``practical like on-ground design,'' translating the general value of agency into specific system functions like structured questions for children. 

\subsection{Designers’ Reflections Corroborated Qualitative Shifts in Agency Reasoning}
\begin{figure}[h]
    \centering
    \includegraphics[width = 0.8\textwidth]{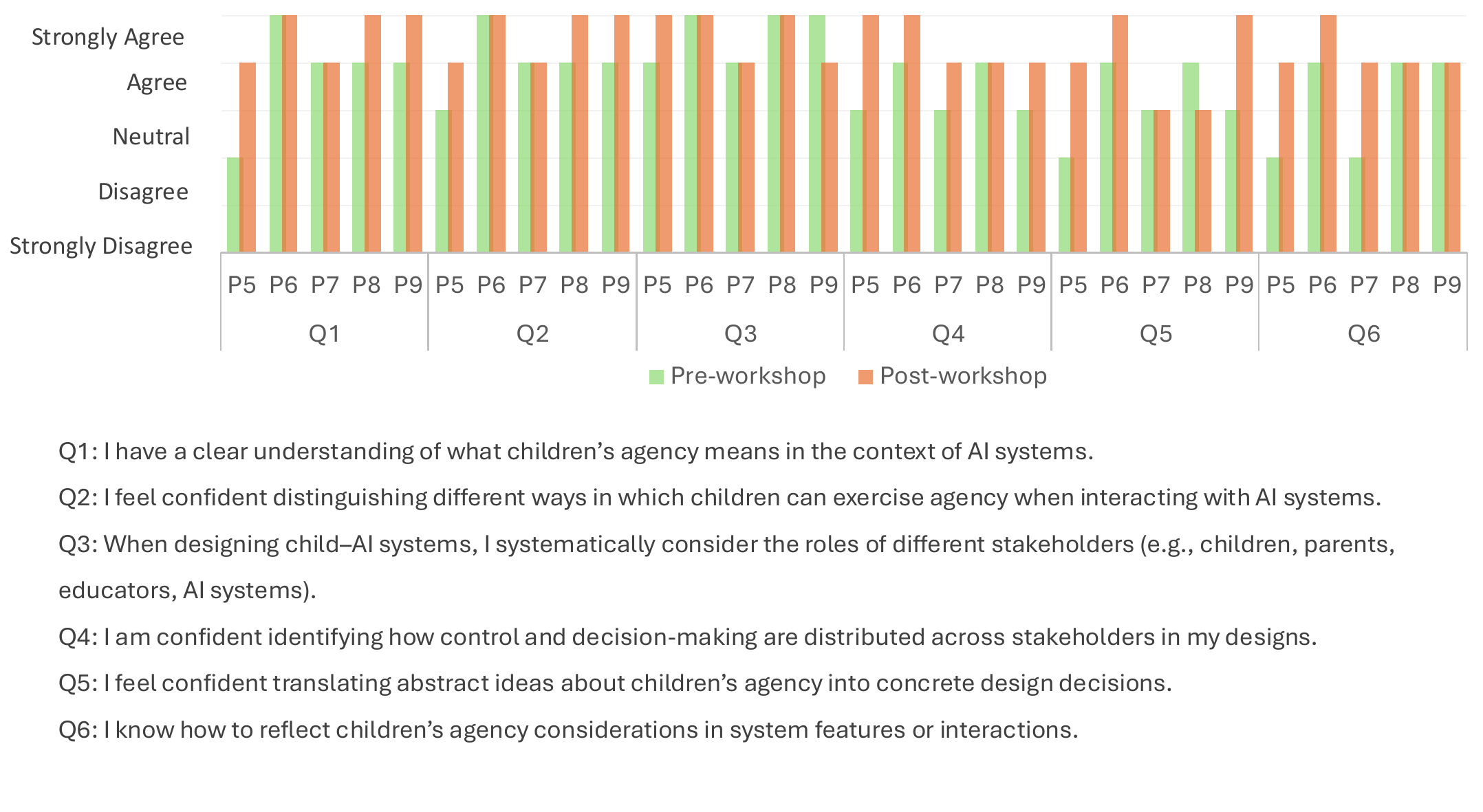}
    \caption{Pre- and post-workshop self-reported agency understanding. Q1 evaluates participants' conceptual understanding of agency. Q2 understands stakeholder awareness. Q3--Q6 estimate concept translation to design. }
    \Description{Pre- and post-workshop self-reported agency understanding. Q1 evaluates participants' conceptual understanding of agency. Q2 understands stakeholder awareness. Q3--Q6 estimate concept translation to design.}
    \label{fig:understanding_of_agency}
\end{figure}

Post-workshop questionnaires reinforced the qualitative findings. Figure~\ref{fig:understanding_of_agency} shows pre- and post-workshop self-reported ratings for each participant across six agency-related questions. Note that we only collected this quantitative data for T1-T3 and post-workshop reflections from T0 was only collected qualitatively. Designers reported \textit{increased awareness} of agency-related trade-offs (Q1) and \textit{greater confidence} in discussing agency explicitly (Q2), while also noting the cognitive effort required to apply the framework consistently. However, designers perception of how they feel about translating agency into concrete design features differed (Q3--Q6), highlighting that designers found operationalisation more challenging than conceptual understanding, consistent with the qualitative observations from the workshops. 

\section{Discussion}
Our exploratory study provided critical evidence of how a conceptual framework that externalises the value of designing for agency could affect designers' reasoning about agency. Rather than introducing new ethical principles, the framework improves designers' awareness and their ability to operationalise the design for agency, shifting them from implicit, experience-based judgments to explicit articulations of agency trade-offs at the component level. This provides critical inputs to the child-centred design community, by presenting an operationalisable conceptual framework prototype and articulating key challenges designers face when reasoning about agency in practice. 

\subsection{Bridging the Gap from Abstract Values to Practical Design}
Despite the small scale of the study, our findings highlight a persistent gap between high-level ethical principles and practical design reasoning. While designers often demonstrate a theoretical understanding of agency (e.g., through psychological or educational backgrounds), this tends to remain implicit and undifferentiated without a structured framework. The CHAI framework functioned as a design thinking scaffold, facilitating designers in explicitly considering different types and levels of agency when translating abstract goals, such as ``protecting the child’s voice'', into system functions. We found the explicit four-stage reasoning process of the framework has been particularly helpful to facilitate the designers to explicitly identify system components related to their agency objectives, annotating and reflecting on their support for agency, surfacing blind spots where adult-centric assumptions previously dominated. This shifted designer' reasoning from intuitive, value-based judgements towards more explicit articulation and justification of agency-related design  decisions. However, our study also showed that a recurring challenge in distinguishing between co-agency and proxy agency. As noted by P7, agency in these contexts is less a simple choice and more about ``who holds the power to make that decision.'' This necessitates treating agency as a relational and negotiable construct while indicates a need to further examine how to better support designers navigating this distinction and balance of power in their designs.

\paragraph{Uncertainty in Allocation of Agency}
A recurring challenge was distinguishing between co-agency and proxy agency in relational contexts.
\begin{itemize}
    \item In T0, a tension emerged between the goal of ``child voice agency'' and the practical reality of parental ``censorship''. Designers realised that if parents ``mark'' or ``evaluate'' every resource the AI provides, the child's individual agency is essentially overwritten by parental proxy agency.
    \item P5 and P6 struggled with whether they were ``determining the agency of the parent or child when logging distress data,'' highlighting the tension between safety and monitoring.
    \item P9 admitted direct uncertainty when categorising their system's primary lens, stating, ``It’s either call agency or proxy agency... I don't know.'' They eventually moved toward co-agency only after reasoning that the system should collect ``parallel information'' from both sides rather than having the parent act solely on the child's behalf.
\end{itemize}
As noted by P7, agency in these contexts is less a simple choice and more about ``who holds the power to make that decision.'' This necessitates treating agency as a relational, negotiable construct.

\paragraph{Agency as a Relational Construct}
These findings point to the importance of treating agency not as a fixed attribute, but as a relational and situational construct that must be actively negotiated in design practice. 
\begin{itemize}
    \item Noted by P7, agency is less about a simple ``choice'' and more about ``who holds the power to make that decision''. 
    \item P9 concluded that using the framework to address these tensions made the process ``not complicated, but complex,'' leading to a more ``refined'' and ``comprehensive'' system that respects both stakeholders.
\end{itemize}

For instance, in T3 designers moved from general intentions to concrete interaction mechanisms (e.g., issue profiles and predefined answer options) that empowered children with limited verbal skills to express preferences within the system.

Furthermore, by requiring designers to annotate and reflect on specific system components, the framework surfaced blind spots where adult-centric assumptions previously dominated. This shifted designer' reasoning from intuitive, value-based judgements towards more explicit articulation and justification of agecy-related design  decisions. 

\subsection{Complexity Empowering Divergent Design Thinking}
Although designers reported that the framework increased the complexity of design processes, they consistently viewed this as a productive constraint that supported more thorough reasoning about user journeys. 
This aligns with the divergent phase of double-diamond design model~\cite{designcouncil2003double}, which encourages explorations before convergence. We observe that the CHAI framework provided explicit reasoning lenses that helped designers explore agency-related trade-offs rather than prematurely simplifying them. As a result, designers engaged more deeply with tensions such as balancing informed consent with high individual agency. For instance, teams considered children’s rights to consent alongside their potential fear of being recorded (T1), or proposed alternative interaction approaches to better reflect the child’s voice (T3). These examples indicate a shift toward more nuanced reasoning about agency within design decisions and a need to further examine how to balance the benefits and potential costs of desgining for agency.


\subsection{Limitations and Future Work}
This study involved only four teams of early-career designers, which limits the generalisability of the findings. While most participants had experiences working with children, they had limited practical design experiences, which may have affected how they engaged with our framework. Future work will expand participant diversity across levels of design expertise, their target child age groups, and AI application domains. Time constraints also limited opportunities to apply the framework at a full system level or to observe downstream implementation.

Participants further identified practical limitations that suggest future development of AI-supported design tools. 
Particularly, distinguishing between low, medium, and high agency often relied on subjective judgment. Designers suggested including a database of case studies and issue profiles to support more consistent reasoning across design contexts. Given the cognitive effort and time required to apply the framework the bottom-up, there is a clear demand for AI-supported scaffolding that can assist designers during early-stage ideation. 

\section{Conclusion}
This paper presents an exploratory study of how a conceptual agency design framework supports designers' reasoning about children’s agency when designing child–AI systems. We show that while designers often hold commitments to children’s well-being and empowerment, agency is rarely articulated explicitly or systematically. 
By introducing the CHAI framework, this study demonstrates how the framework enabled designers to incorporate agency into design reasoning as an explicit, discussable, and configurable concern. 
We argue that designing for children’s agency is fundamentally a reasoning challenge rather than an outcome-oriented problem. Supporting this reasoning requires tools that embrace complexity, make tensions visible, and enable negotiation rather than simplification. By framing agency as relational, distributed, and situational, this work contributes an empirical account of how designers think with agency and offers directions for future research on agency-oriented design support for child–AI systems.



\bibliographystyle{ACM-Reference-Format}
\bibliography{sample-base,related,allbib}
\end{document}